\documentstyle[12pt,preprint,aps,epsfig]{revtex}
\tighten
\newcommand{\beq}{\begin{eqnarray}}
\newcommand{\eeq}{\end{eqnarray}}
\newcommand{\bsga}{ b \to s \gamma }

\newcommand{\calb}{ {\cal B}}

\newcommand{\mhp}{ M_{H^{+}}}

\def\etap{\eta^{\prime}}
\def\etapp{\eta^{(')}}

\def\lsim{ {\ \lower-1.2pt\vbox{\hbox{\rlap{$<$}\lower5pt\vbox{\hbox{$\sim$}}}}\ } }

\newcommand{\tab}[1]{Table \ref{#1}}

\newcommand{\non}{\nonumber\\ }
\title{\bf{ $B \to \pi^+ \pi^-,\;  K \pi, \; K \etap$ decays and new physics
effects in the general two-Higgs-doublet model: an update } }
\author{ Zhenjun Xiao\thanks{E-mail: zjxiao@email.njnu.edu.cn},\\
{\small  Department of Physics, Nanjing Normal University, Nanjing,
 210097, People's Republic of China }\\
Kuang-Ta Chao and Chong Sheng Li\\ {\small Department of Physics, Peking
University, Beijing, 100871, People's Republic of China} }
\date{\today}
\begin{document}
\maketitle
\begin{abstract}
In this paper, we reexamine the new physics contributions to seven well measured
$B \to PP$ decays  in the standard model (SM) and the general two-Higgs-doublet
model (model III) and compare the
theoretical predictions with the new data. Within the considered parameter space we found that:
(a) the measurements of the branching ratios for $B\to \pi^+ \pi^-,  K^- \pi^+$ and $K^0 \pi^+$
lead to a strong constraint on the form factor $F_0^{B\pi}(0)$: $F_0^{B\pi}(0)=0.24 \pm 0.03$;
(b) the new physics enhancements  to the penguin-dominated $B \to K \pi$ and $K \etap$
decays are significant in size, $\sim (40-65)\%$ $w.r.t$ the SM predictions,
and play an important role in restoring the consistency between the data and the theory.
\end{abstract}


\vspace{0.5cm} \noindent

\newpage

As is well known, one of the main objectives of B experiments is to probe for the
possible effects of new physics beyond the standard model (SM).
Precision measurements of the B meson system can provide an insight into very
high energy scales via the indirect loop effects of the new physics\cite{slac504,xiao20}.

Up to now,  CLEO, BaBar and Belle Collaboration \cite{cleo2000,babar2001,belle2001}
have observed eighteen two-body charmless hadronic  $B_{u,d}$ meson decay modes. Seven well
measured $B \to P P$ ( here $P$ refers to the light pseudo-scalar mesons ) decays are
\beq
B\to \pi^\pm \pi^\mp,\; K \pi,\;  K \etap\; . \label{eq:bpp7}
\eeq
These decay modes are closely related through the isospin symmetry and the $SU(3)$
flavor symmetry, phenomenologically
very interesting due to their key role in extracting the unitary angles $\alpha$
and $\gamma$,  and the appearance of the so-called $\etap K$
puzzle: the observed $B \to K\etap$ decay rates\cite{cleo2000,babar2001,belle2001}
are much larger than what expected in the standard model
based on the effective Hamiltonian with generalized factorization (GF) approach
\cite{bbl96,bsw87,ali9804,chen99}.
To accommodate the data, one may need an additional contribution unique to the
$\etap $ meson in the framework of the SM \cite{as97,chao97,du98,hz98},
or enhancements from new physics models beyond the SM \cite{x10326,np}.

In a previous paper\cite{x10326}, we considered the second possibility and calculated
the new physics effects on the two-body  charmless hadronic
B meson decays in the general two-Higgs-doublet models (2HDM's) \cite{atwood97}.
In this paper, we focus on seven well measured $B \to P P$ decay modes and compare
the theoretical predictions with the newest data.
We still use the low-energy effective Hamiltonian with GF approach \cite{ali9804,chen99}
to calculate the new physics contributions. For recent studies of $B \to P P$ decays
in the SM with the QCD factorization (BBNS) and perturbative QCD (PQCD) approaches
\cite{bbns99,pqcd}, one can see the papers
\cite{li2001,du2000,yang00,du2002} and references therein.

For the  inclusive three-body decays $b \to s \bar{q} q $
with $q\in \{u,d,s \}$ the effective Hamiltonian can be written as \cite{bbl96,ali9804},
\begin{equation}
{\cal H}_{eff}(\Delta B=1) = \frac{G_F}{\sqrt{2}} \left \{
\sum_{j=1}^2 C_j \left ( V_{ub}V_{us}^* Q_j^u  + V_{cb}V_{cs}^*
Q_j^c \right ) - V_{tb}V_{ts}^* \left [ \sum_{j=3}^{10}  C_j Q_j +
C_{g} Q_{g} \right ] \right \}~. \label{heff2}
 \end{equation}
The explicit expressions for all operators can be found easily for example in Ref.\cite{ali9804}.
For $b \to d \bar{q} q$ decays, one simply makes the replacement $s \to d$.
Following Ref.\cite{ali9804}, we also neglect the effects
of the electromagnetic penguin operator $Q_{7\gamma}$, the weak annihilation and
exchange diagrams. Within the SM and at scale $M_W$,
the Wilson coefficients $C_1(M_W), \cdots, C_{10}(M_W)$ at next-to-leading logarithmic
order {NLO) and $C_{g}(M_W)$ at leading logarithmic order (LO) have been given
for example in Ref.\cite{bbl96}.

In a recent paper \cite{chao99}, Chao {\it et al.} studied the decay $\bsga$ in
model III (the third type of 2HDM's) by assuming that only the couplings $\lambda_{tt}=|\lambda_{tt}|
e^{i\theta_t}$ and $\lambda_{bb}=|\lambda_{bb}|e^{i\theta_b}$ are non-zero.
They found that the constraint on $\mhp$ imposed by the CLEO
data of $\bsga$ can be greatly relaxed by considering the phase
effects of $\lambda_{tt}$ and $\lambda_{bb}$. From the studies of
Refs.\cite{xiao20,chao99}, we know that for model III the parameter space
\beq
&& \lambda_{ij}=0, \ \ for \ \ ij\neq tt,\ \ or \ \  bb, \nonumber\\
&& |\lambda_{tt}|= 0.3,\ \ |\lambda_{bb}|=35,\ \
\theta=(0^0 - 30^0),\ \ \mhp=(200 \pm 100 ){\rm GeV}, \label{eq:lm3}
\eeq
are allowed by the available data, where $\theta=\theta_{bb}-\theta_{tt}$.
In this paper, we calculate the new physics contributions to seven B meson decay
modes in the Chao-Cheung-Keung (CCK) scenario of model III \cite{chao99}.

Following the same procedure as in the SM, it is straightforward to calculate the new
$\gamma$-, $Z^0$- and gluonic penguin diagrams induced by the exchanges of charged-Higgs
bosons appeared  in model III (for details of the calculations, see Ref.\cite{x10326}).
In the NDR scheme, by using the input parameters as given in Eqs.(\ref{eq:lm3}) and
setting $\mu=2.5$ GeV,  we find that:
\beq
&& C_1 =  1.1245,\ \   C_2 = - 0.2662, \ \ C_3  =  0.0186,\ \   C_4 = -0.0458,\non
&& C_5 =  0.0113, \ \ C_6  =  -0.0587,\ \  C_7  =  0.0006, \ \  C_8  = 0.0007,\non
&& C_9 =  -0.0096,\ \ C_{10} = 0.0026,  C_g^{eff}  =  0.3364 \label{eq:cgmb2}
\eeq
where $C_g^{eff} = C_{8G}+ C_5$.

For the $B \to PP$ decay modes considered here, the decay amplitudes as given in
Ref.\cite{ali9804} will be used without further discussion about details. We focus on
estimating the new physics effects on those seven well measured decay modes.
In the NDR scheme and for $SU(3)_C$, the effective Wilson coefficients
can be written as \cite{chen99}
\beq
C_i^{eff} &=& \left [ 1 + \frac{\alpha_s}{4\pi} \, \left( \hat{r}_V^T +
 \gamma_{V}^T \log \frac{m_b}{\mu}\right) \right ]_{ij} \, C_j
 +\frac{\alpha_s}{24\pi} \, A_i' \left (C_t + C_p + C_g \right)
+ \frac{\alpha_{ew}}{8\pi}\, B_i' C_e ~, \label{eq:wceff}
\eeq
where $A_i'=(0,0,-1,3,-1,3,0,0,0,0)^T$, $B_i'=(0,0,0,0,0,0,1,0,1,0)^T$, the
matrices  $\hat{r}_V$ and $\gamma_V$ contain the process-independent
contributions from the vertex diagrams\cite{chen99,x10326}.
The function $C_t$, $C_p$, and $C_g$ describe the contributions arising
from the penguin diagrams of the current-current
$Q_{1,2}$, the QCD operators $Q_3$-$Q_6$, and the tree-level diagram of the
magnetic dipole operator $Q_{8G}$, respectively. The explicit expressions of
the functions $C_t$, $C_p$, and $C_g$ can be found for example in Ref.\cite{x10326}.

In the generalized factorization approach, the effective Wilson coefficients $C_i^{eff}$
will appear in the decay amplitudes in the combinations,
\begin{equation}
a_{2i-1}\equiv C_{2i-1}^{eff} +\frac{{C}_{2i}^{eff}}{N_c^{eff}}, \ \
a_{2i}\equiv C_{2i}^{eff}     +\frac{{C}_{2i-1}^{eff}}{N_c^{eff}}, \ \ \
(i=1,\ldots,5) \label{eq:ai}
\end{equation}
where the effective number of colors $N_c^{eff}$ is treated as a free parameter
varying in the range of $2 \leq N_c^{eff} \leq \infty$, in order to model the
non-factorizable contribution to the hadronic matrix elements.

In the B rest frame, the branching ratios ${\cal B}(B \to PP)$ can be written as
\begin{equation}
{\cal B}(B \to X Y )=  \tau_B \frac{|p|}{8\pi M_B^2}
|M(B\to XY)|^2~,\label{eq:brbpp}
\end{equation}
where $\tau_B=1.653\,  ps$ and $1.548\, ps$ for $B=B_u^-$ and $B_d^0$\cite{pdg2000}, respectively.
$p_B$ is the four-momentum of the B meson, $M_B=5.279$ GeV is the mass of $B_u$ or $B_d$ meson.
In the numerical calculations we use the same input parameters (the masses, gauge couplings,
decay constants, form factors, {\it etc.}) as in Ref.\cite{x10326}.
Particularly, the elements of CKM matrix in the Wolfenstein parametrization
are: $A=0.81,\;  \lambda=0.2205,\;  \rho=0.12, \; \eta=0.34$, which corresponds to
$\gamma=71^\circ$ and $\sin{2\beta}=0.79$ favored by the global fit
and the new measurements\cite{prl87-09}.

For the seven well measured $B \to PP$ decay modes, currently available measurements as
reported by CLEO, BaBar and Belle Collaborations \cite{cleo2000,babar2001,belle2001} and their
averages are listed in Table \ref{exp}. The data have been changed greatly when compared
with those in year 2000:
\begin{itemize}
\item
For ratio $\calb(B \to \pi^+ \pi^-)$, the new BaBar result is $(4.1 \pm 1.0 \pm 0.7)\times 10^{-6}$
instead of the old $(9.3 ^{+2.8\; +1.2 }_{-2.1\; -1.4} )\times 10^{-6}$.
The average therefore decreased to $(4.4 \pm 0.9 )\times 10^{-6}$.

\item
The BaBar measurement of $\calb (B \to K^0 \pi^0$) is only $\sim 8 \times 10^{-6}$,
the average therefore becomes much smaller than two years ago.

\item
For the ratio $\calb (B \to K^0 \etap)$, both the BaBar and Belle result are
much smaller than CLEO's measurement, the average is only $(56 \pm 10)\times
10^{-6}$ and clearly smaller than the branching ratio of $B \to K^+ \etap$ decay.
\end{itemize}

In the SM and GF approach, the decay $B \to \pi^+ \pi^-$ is very simple and receives contributions
from the dominated tree diagram, the QCD and electroweak penguin diagrams, and
depend on one form factor $F_0^{B\pi}(0)$ only, as can be seen from the decay
amplitude \cite{ali9804}:
\beq
{\cal M}(\bar{B}^0 \to \pi^+ \pi^-)&=& -i \frac{G_F}{\sqrt{2}} f_\pi
F_0^{B\pi}(m_\pi^2)\left ( m_B^2- m_\pi^2 \right ) \non
&& \times \left \{  V_{ub} V_{ud}^* a_1
- V_{tb} V_{td}^* \left [
a_4 + a_{10} + 2 \left ( a_6 + a_8 \right )\frac{m_\pi^2}{(m_b -m_u)(m_u + m_d)}\right ]\right \}
\label{eq:bpipi}
\eeq
The new physics contribution to this decay in the model III is only $2.5\%$ and thus can be
neglected. In the SM, there is no other contribution to this decay mode and therefore it seems
to be a clean decay mode to determine the form factor $F_0^{B\pi}(0)$.
For $\gamma\approx 71^\circ$ as indicated by the global fit, the interference
between the tree and penguin-diagrams is constructive, and the measured branching
ratio of this decay leads to a small $F_0^{B\pi}(0)$
\beq
F_0^{B\pi}(0)=0.21 \pm 0.03. \label{eq:f00}
\eeq
Which is clearly smaller than the values from Lattice-QCD  or Light-cone QCD sum
rules: $F_0^{B\pi}(0)=0.30 \pm 0.04$, or $0.28 \pm 0.05$  as given in Ref.\cite{ball98}
and Ref.\cite{khod}, respectively. For $F_0^{B\pi}(0)=0.3$ and $\gamma=71^\circ$, however,
the SM prediction for the branching ratio is $(7.5 -10.7)\times 10^{-6}$ for $N_c^{eff}=2-\infty$
in the GF approach and about $9\times 10^{-6}$  in the BBNS approach\cite{muta00},
which is clearly too large to be consistent with the data. We thus believe that
the form factor $F_0^{B\pi}(0)$ should be apparently smaller than $0.3$.

On the other hand, the QCD-penguin-dominated $B \to K^- \pi^+$ and $K^0 \pi^+$ decays
also depend on the form factor $F_0^{B\pi}(0)$, as can be seen from the decay amplitudes
as given in Ref.\cite{ali9804}. But one should be very careful to extract the form factor
$F_0^{B\pi}(0)$ from these two decay modes because of  (a)
the neglected rescattering and other non-factorized
contributions to these two decays may be large; and (b) the new physics contributions in the
model III are also large, $\sim 50\%$ respect to the SM predictions. In the GF approach, the
measured branching ratios of these two decays prefer a larger $F_0^{B\pi}(0)$:
$F_0^{B\pi}(0)=0.30 \pm 0.03$ if the new physics contributions to these two decay modes
are not included. But
\beq
F_0^{B\pi}(0)=0.24 \pm 0.03. \label{eq:f03}
\eeq
if the measured branching ratios of $B \to \pi^+ \pi^-, K^- \pi^+, K^0 \pi^+$
decays  and the new physics enhancements are all taken into account.
Although the central value of the form factor in Eq.(\ref{eq:f03}) is still smaller
than $F_0^{B\pi}(0)=0.28 \pm 0.05$ as given in Refs.\cite{ball98,khod}, they are
compatible within errors. In the numerical calculations, we will use
$F_0^{B\pi}(0)=0.24 \pm 0.03$.

Furthermore, the form factor $F_0^{BK}(0)$ cannot deviate too much from $F_0^{B\pi}(0)$,
otherwise the $SU(3)$ flavor symmetry will be broken badly. As indicated by the data
and theoretical considerations, it is a good approximation to take
$F_0^{BK}(0)/F_0^{B\pi}(0) = f_k /f_\pi$ as a measure of $SU(3)$ symmetry breaking.
We then find that
\beq
F_0^{BK}(0)=0.29 \pm 0.04. \label{eq:f04}
\eeq
for $F_0^{B\pi}(0)=0.24 \pm 0.03$.

In the GF approach, the QCD-penguin-dominated $ B \to K \etap$ decays depend on the form
factors $F_0^{BK}(0)$ and $F_0^{B\etap}(0)$,
\beq
F_0^{B\etap}(0)= F_0^{B\pi}(0) \left [ \frac{\sin
\theta_8}{\sqrt{6}} + \frac{\cos \theta_0}{\sqrt{3}} \right ]= 0.10 \pm 0.02, \label{eq:f05}
\eeq
for $F_0^{B\pi}(0)=0.24 \pm 0.03$, $\theta_0=-9.1^\circ$ and $\theta_8=-22.2^\circ$ in the
two-angle mixing scheme\cite{fk98}.

In \tab{bpp1}, we show the theoretical predictions for the branching ratios of seven studied decay
modes. We use the form factors as given in Eqs.(\ref{eq:f03},\ref{eq:f04},\ref{eq:f05}), and
keep all other input parameters the same as those used in Ref.\cite{x10326}.
The branching ratios collected in \tab{bpp1} are the averages of
the corresponding $B$ and anti-$B$ meson decay rates. The ratio $\delta
{\cal  B}$  describes the new physics correction on the decay rates and is defined as
\beq
\delta {\cal  B} (B \to XY) = \frac{{\cal  B}(B \to XY)^{III}
-{\cal  B}(B \to XY)^{SM}}{{\cal  B}(B \to XY)^{SM}} \label{eq:dbr}
\eeq

From \tab{exp} and \tab{bpp1}, we find that
\begin{itemize}
\item
By using $F_0^{B\pi}(0)=0.24 \pm 0.03$, our predictions for $\calb (B \to \pi^+
\pi^-)$ in both the SM and model III are in agreement with the data.

\item
For $B \to K \pi$ decays, the SM predictions seem smaller than the
measurements. The results under the BBNS approach are similar \cite{du2002}.
For the $B \to K^0 \pi^0$ decay, specifically, the SM
prediction is about half of the measured decay rate, and the new physics
enhancement is essential for the theoretical prediction to become
consistent with the data.

\item
By taking into account the uncertainties of those
input parameters as given explicitly in Ref.\cite{x10326}, we find numerically that
\beq
{\cal B}(B \to K^+ \etap )\approx {\cal B}(B \to K^0 \etap ) = \left \{\begin{array}{ll}
( 10-40 )\times 10^{-6} & {\rm in \ \ SM }~, \\
( 17-57 )\times 10^{-6} & {\rm in  \ \ Model \ \ III}~.  \\
\end{array} \right. \label{eq:betap1}
\eeq
where the uncertainties of those input parameters have been considered.
It is easy to see that the SM predictions in the GF approach is about half of
the measured value. In BBNS approach, the theoretical predictions for the corresponding
branching ratios are also much smaller than the experimental data\cite{du2002}.
The new physics enhancement can boost the theoretical predictions close to the lower part of the
measured values, but still leaves a moderate space for additional contributions.
\end{itemize}

Since 1997, the unexpected large $\etap$ production has been widely discussed in the literature
\cite{as97,chao97,du98,hz98,x10326,np,du2000,yang00,du2002,muta00}. For the sake of completeness,
we here make a brief comment on some typical interpretations in the framework of the SM.
\begin{itemize}
\item
Atwood and Soni \cite{as97} gave arguments for the need of enhanced $b \to s
g^*$ decays followed by $g^* \to \etap g $ via the QCD gluon anomaly.
Taking a constant $gg\etap$ vertex form factor $H(0,0,m_{\etap}^2)$, the observed
large branching ratio $B \to \etap X_s$ can be achieved. But as pointed by
Hou and Tseng\cite{as97}, Kagan and Petrov\cite{np}, if one considers the
running of $\alpha_s$, and the $m^2_{\etap}/(q^2 -m^2_{\etap})$ dependence of the
$gg\etap$ coupling, the result presented in \cite{as97} will be reduced
greatly.

\item
Halperin and Zhitnitsky \cite{hz98} argued that the dominant contribution to
$\etap$ production is due to the Cabibbo favored $b \to (\bar{c}c)_1 s$ process
followed by the transition $(\bar{c}c)_1 \to \etap$, i.e. the "intrinsic charm"
component of $\etap$. But according to the explicit calculations in
Refs.\cite{ali9804,chen99}, this mechanism can not give a good explanation for the
measured $B \to K \etap$ decay rates.

\item
Yuan and Chao \cite{chao97} argued that the inclusive $\etap$ production in B
decays may dominantly come from the Cabibbo favored $b \to (\bar{c}c)_8 s$ process
where $\bar{c}c$ pair is in a color-octet configuration, and followed by the
nonperturbative transition $(\bar{c}c)_8 \to \etap X$.

\item
The authors of Ref.\cite{du98}
proposed di-gluon fusion mechanism. It seems that this mechanism could enhance
$\etap$ production. But because of our ignorance about the form factor of $g^* g^*
\etap$ vertex, there are large uncertainties in calculation.

\item
Recently, M.Z.Yang and Y.D.Yang reconsidered the di-gluon mechanism and gave a
calculation for $B \to \etapp P$ in the BBNS approach. They computed the vertex
$g^* g^* \etap$ in perturbative QCD, and found that the branching ratios of
$B \to K \etap$
are really enhanced and in agreement with the data. But, as pointed in
Ref.\cite{du2002}, the consistency of Yang's perturbative calculation is
questionable due to the end point behavior.

\end{itemize}
From above discussions, one can understand that it is still an open question for us
to interpret the large branching ratios of $B \to K \etap$ decays.
Further investigations for various possible mechanisms are welcome.

Furthermore, because of the isospin symmetry between the $u$ and $d$ quarks, the decays
$B \to K^+ \etap$ and $K^0 \etap$ should theoretically have similar branching ratios.
The known new mechanisms in the SM or the new physics models also
contribute to these two decay modes in very similar way. The measurements of the BaBar
and Belle collaboration, however,  show a clear difference between these two decay rates.
We do not know how to interpret this difference. One may need something new to resolve
this problem if it is confirmed by the forthcoming data.

In short, we  reexamined the branching fractions of seven well measured $B \to P P$ decay
modes in the SM and model III, and compared  the theoretical predictions with the new data.
Within the considered parameter space we found that:
(a) the measurements of the branching ratios for $B\to \pi^+ \pi^-,  K^- \pi^+$ and $K^0 \pi^+$
lead to a strong constraint on the form factor $F_0^{B\pi}(0)$: $F_0^{B\pi}(0)=0.24 \pm 0.03$;
(b) the new physics enhancements  to the penguin-dominated $B \to K \pi$ and $K \etap$
decays are significant in size, $\sim (40-65)\%$ $w.r.t$ the SM predictions,
and play an important role in restoring the consistency between the data and the theory.

\section*{ACKNOWLEDGMENTS}
K.T.~Chao and C.S.~Li acknowledge the support by the National Natural
Science Foundation of China,  the State Commission of Science
and technology of China. Z.J.~Xiao acknowledges the support by the National
Natural Science Foundation of China under Grant No.10075013,
and by the Research Foundation of Nanjing Normal University under Grant No.2001WLXXGQA916.

\newpage

\begin{table}[t]
\begin{center}
\caption{Measurements of the branching ratio ${\cal B}(B\to PP)$ (in units of $10^{-6}$)
as reported by CLEO, BaBar and Belle Collaborations. The last column lists the world
average.}
\label{exp}
\vspace{0.2cm}
\begin{tabular} {|l|l|l|l|l|}  \hline
Decay Mode & CLEO & BaBar & Belle & Average  \\ \hline
$B^0 \to \pi^+ \pi^-$         &$4.3^{+1.6}_{-1.5}\pm 0.5$ & $ 4.1 \pm 1.0 \pm 0.7$
 &$5.6^{+2.3}_{-2.0}\pm 0.4 $ &$4.4 \pm 0.9$  \\ \hline
$B^0 \to K^+ \pi^-$           &$17.2^{+2.5}_{-2.4}\pm 1.2$&$16.7 \pm 1.6 \pm 1.3$
    &$19.3 ^{+3.4 +1.5}_{-3.2 -0.6}$& $17.3\pm 1.5$ \\
$B^+ \to K^+ \pi^0$           &$11.6^{+3.0 +1.4}_{-2.7 - 1.3}$&$10.8^{+2.1}_{-1.9}\pm 1.0$
    & $16.3^{+3.5 +1.6}_{-3.3 -1.8} $&$12.1\pm 1.7$  \\
 $B^+ \to K^0 \pi^+$           &$18.2 ^{+4.6}_{-4.0}\pm 1.6$&$18.2 ^{+3.3}_{-3.0} \pm 2.0$
    &$13.7^{+5.7 +1.9}_{-4.8 -1.8} $& $17.4 \pm 2.6$ \\
$B^0 \to K^0 \pi^0$           &$14.6 ^{+5.9 +2.4}_{-5.1 -3.3}$&$8.2 ^{+3.1}_{-2.7} \pm 1.2$
    &$16.0 ^{+7.2 +2.5}_{-5.9 -2.7}$&$10.7 \pm 2.7$  \\ \hline
$ B^+ \to  K^+ \eta^\prime$   & $80^{+10}_{-9} \pm 7$& $70 \pm 8 \pm 5$
    & $79^{+12}_{-11} \pm 9 $ & $75 \pm 7 $  \\
$B^0 \to K^0 \eta^\prime$     &$89^{+18}_{-16} \pm 9$& $42^{+13}_{-11} \pm 4$
    & $55^{+19 }_{-16} \pm 8 $ & $56\pm 10$ \\ \hline
\end{tabular}\end{center}
\end{table}

\begin{table}
\begin{center}
\caption{Branching ratios (in units of $10^{-6}$) of seven  $B \to PP$ decay
modes in the SM and Model III  by using $F_0^{B\pi}(0)=0.21$ (the first
entries), $0.24$ (the second entries),  $0.27$ (the third entries),
and assuming $N_c^{eff}=2-\infty$,
$\theta=0^\circ$ and $\mhp=200$ GeV. The last column lists the world average data. }
\label{bpp1}
\vspace{0.2cm}
\begin{tabular} {l ccc ccc c l} \hline
 &  \multicolumn{3}{c}{SM }&
\multicolumn{3}{c}{Model III}& $\delta {\cal  B} \; [\%]$ & Data   \\
\cline{2-9}
Decay Mode & $2$& $3$ & $\infty$ & $2$& $3$ & $\infty$&$N_c^{eff}=3$& \\ \hline
$B^0 \to \pi^+ \pi^-$       & $3.66$ &$4.16$&$5.24$&$3.75$ &$4.26$&$5.37$&$2.5 $&$4.4 \pm 0.9$ \\
                            & $4.78$ &$5.43$&$6.85$&$4.90$ &$5.56$&$7.01$&$2.5 $& \\
                            & $6.05$ &$6.87$&$8.66$&$6.20$ &$7.04$&$8.87$&$2.5 $& \\ \hline
$B^+ \to K^+ \pi^0$         & $4.93$ &$5.53$&$6.85$&$7.12$ &$8.01$&$9.98$&$44.9$&$12.1\pm 1.7$ \\
                            & $6.44$ &$7.22$&$8.95$&$9.29$ &$10.5$&$13.0$&$44.9$&\\
                            & $8.15$ &$9.14$&$11.3$&$11.8$ &$13.2$&$16.5$&$44.9 $& \\ \hline
$B^0 \to K^+ \pi^-$         & $7.18$ &$7.96$&$9.66$&$10.8$ &$12.0$&$14.7$&$51.3$&$17.3\pm 1.3$ \\
                            & $9.37$ &$10.4$&$12.6$&$14.1$ &$15.7$&$19.2$&$51.3$&\\
                            & $11.9$ &$13.2$&$16.0$&$17.9$ &$19.9$&$24.3$&$51.3 $&\\ \hline
$B^+ \to K^0 \pi^+$         & $8.12$ &$9.46$&$12.5$&$12.1$ &$14.1$&$18.4$&$48.9$&$17.4\pm 2.6$\\
                            & $10.6$ &$12.4$&$16.3$&$15.9$ &$18.4$&$24.1$&$48.9$&\\
                            & $13.4$ &$15.6$&$20.6$&$20.1$ &$23.3$&$30.5$&$48.9 $& \\
                            \hline
$B^0 \to K^0 \pi^0$         & $2.90$ &$3.32$&$4.25$&$4.56$ &$5.21$&$6.66$&$56.9$&$10.7\pm 2.7$ \\
                            & $3.80$ &$4.34$&$5.56$&$5.96$ &$6.80$&$8.69$&$56.9$&\\
                            & $4.81$ &$5.49$&$7.03$&$7.54$ &$8.61$&$11.0$&$56.9$& \\
                            \hline
$ B^+ \to  K^+ \eta^\prime$ & $9.69$ &$12.2$&$18.2$&$16.3$ &$20.1$&$29.0$&$64.7$&$75  \pm 7  $\\
                            & $12.7$ &$16.0$&$23.8$&$21.3$ &$26.3$&$37.9$&$64.7$&\\
                            & $16.0$ &$20.2$&$30.5$&$27.0$ &$33.3$&$47.9$&$64.7 $& \\
                            \hline
$B^0 \to K^0 \eta^\prime$   & $9.33$ &$12.0$&$18.3$&$15.6$ &$19.5$&$30.0$&$62.8$&$56  \pm 10 $ \\
                            & $12.2$ &$15.6$&$23.9$&$20.4$ &$25.5$&$37.4$&$62.8$&\\
                            & $15.4$ &$19.8$&$30.2$&$25.8$ &$32.2$&$47.3$&$62.8 $& \\
                            \hline
\end{tabular}\end{center} \end{table}

\end{document}